\documentclass[prd,aps,twocolumn,nofootinbib,preprintnumbers,superscriptaddress,preprintnumbers,balancelastpage,longbibliography]{revtex4-2}
\usepackage[english]{babel}

\usepackage{amsmath,mathtools,physics,xfrac}
\usepackage[normalem]{ulem}
\usepackage{graphicx}
\usepackage{afterpage}
\usepackage{float}
\usepackage{subfigure}
\usepackage{rotating}
\usepackage{multirow}
\usepackage{tabularx}
\usepackage{booktabs}
\usepackage{fancyhdr}
\usepackage[utf8]{inputenc}
\usepackage{theorem}
\usepackage{moreverb}
\usepackage{euscript}
\usepackage{psfrag}
\usepackage{slashed}
\usepackage{mathtools}
\usepackage{makecell}
\usepackage{adjustbox}
\usepackage{dcolumn}
\usepackage{bm}
\usepackage[dvipsnames]{xcolor}
\usepackage{graphics}
\usepackage{hyperref}
\usepackage{lettrine}

\input Zallman.fd

\LettrineTextFont{\itshape}
\hypersetup{
     colorlinks   = true,
     citecolor    = Cyan,
     urlcolor     = Cyan,
     linkcolor    = Cyan
}

\usepackage{color,xcolor}

\definecolor{darkblue}{rgb}{0.0,0.0,0.75}
\definecolor{darkred}{rgb}{0.6,0.0,0}
\definecolor{darkgreen}{rgb}{0.0,0.6,0.}

\newcommand{\K}{\,\mathrm{K}}
\newcommand{\W}{\,\mathrm{W}}

\newcommand{\mum}{\,\mu\mathrm{m}}

\newcommand{\Wmum}{\,\mathrm{W}\mu\mathrm{m}^{-3}}

\newcommand{\GeV}{\,\mathrm{GeV}}

\newcommand{\mueV}{\,\mathrm{\mu eV}}

\newcommand{\cmsq}{\,\mathrm{cm^{2}}}

\newcommand{\mchi}{m_\chi}
\newcommand{\nchi}{n_\chi}
\newcommand{\bq}{\boldsymbol{q}}
\newcommand{\oq}{\omega_{\boldsymbol{q}}}
\newcommand{\bv}{\boldsymbol{v}}
\newcommand{\fmed}{F_\mathrm{med}}
\newcommand{\sigman}{\sigma_{\chi N}}

\newcommand{\nqp}{n_\mathrm{qp}}

\newcommand\redsout{\bgroup\markoverwith{\textcolor{red}{\rule[0.5ex]{2pt}{0.4pt}}}\ULon}

\begin{document}

\preprint{SLAC-PUB-17769}

\title{Transmon Qubit Constraints on Dark Matter-Nucleon Scattering}

\author{Anirban~Das}
\thanks{\href{mailto:anirbandas@snu.ac.kr}{anirbandas@snu.ac.kr}; \href{http://orcid.org/0000-0002-7880-9454}{0000-0002-7880-9454}
}
\affiliation{Center for Theoretical Physics, Department of Physics \& Astronomy, Seoul National University, Seoul 08826, South Korea
}

\author{Noah~Kurinsky}
\thanks{\href{mailto:kurinsky@slac.stanford.edu}{kurinsky@slac.stanford.edu}; \href{http://orcid.org/0000-0002-5872-519X}{0000-0002-5872-519X}
}
\affiliation{SLAC National Accelerator Laboratory, 2575 Sand Hill Rd, Menlo Park, CA 94025, USA}
\affiliation{Kavli Institute for Particle Astrophysics and Cosmology, Stanford University, Stanford, CA 94305, USA}

\author{Rebecca~K.~Leane}
\thanks{\href{mailto:rleane@slac.stanford.edu}{rleane@slac.stanford.edu}; \href{http://orcid.org/0000-0002-1287-8780}{0000-0002-1287-8780}
}
\affiliation{SLAC National Accelerator Laboratory, 2575 Sand Hill Rd, Menlo Park, CA 94025, USA}
\affiliation{Kavli Institute for Particle Astrophysics and Cosmology, Stanford University, Stanford, CA 94305, USA}

\date{June 25, 2024}

\begin{abstract}
We recently pointed out that power measurements of single quasiparticle devices can be used to detect dark matter. These devices have the lowest known energy thresholds, far surpassing standard direct detection experiments, requiring energy deposition above only about an meV. We calculate dark matter induced quasiparticle densities in transmon qubits, and use the latest transmon qubit measurements that provide one of the strongest existing lab-based bounds on dark matter-nucleon scattering below about 100 MeV. We strongly constrain sub-component dark matter, using both a dark matter population thermalized in the Earth as well as the dark matter wind from the Galactic halo. We demonstrate future potential sensitivities using devices with low quasiparticle densities.
\end{abstract}

\maketitle

\lettrine{D}{irect detection} experiments are currently the main detection effort for dark matter scattering on Earth. As the dark matter wind impinges on Earth's surface, dark matter particles can deposit energy inside direct detection experiments, which search for anomalous recoils of their Standard Model (SM) targets. This provides a robust lab-based test of the dark matter-SM scattering cross section. However, direct detection experiments are limited to dark matter masses above about a GeV for analyses assuming nuclear recoils~\cite{LUX-ZEPLIN:2022qhg}, or MeV-scale masses when exploiting the Migdal effect~\cite{Vergados:2005dpd, Moustakidis:2005gx, Bernabei:2007jz, Ibe:2017yqa,XENON:2019zpr,PandaX:2022xqx} or electron recoils~\cite{Tiffenberg:2017aac,SENSEI:2020dpa,XENON:2019gfn,DAMIC-M:2023gxo}.

To date, no definitive dark matter induced recoil signal has been detected, motivating a push for new approaches to search for dark matter scattering outside the known sensitivity of direct detection experiments~\cite{Hochberg:2022apz}. To probe lighter dark matter, or dark matter with less energy than that coming from the dark matter wind, lower threshold devices are required. New proposals have been made in recent years on multiple fronts, with new detection possibilities with superconductors~\cite{Hochberg:2015pha, Hochberg:2015fth, Hochberg:2021pkt, Hochberg:2021ymx,Hochberg:2019cyy,Chiles:2021gxk,Chen:2022quj,Chen:2023swh}, superfluids~\cite{Schutz:2016tid, Knapen:2016cue, Caputo:2019cyg}, polar crystals~\cite{Griffin:2018bjn, Knapen:2017ekk, Cox:2019cod}, doped semiconductors~\cite{Du:2022dxf}, topological materials~\cite{Sanchez-Martinez:2019bac}, Dirac materials~\cite{Hochberg:2017wce, Geilhufe:2018gry, Geilhufe:2019ndy, Coskuner:2019odd,Das:2023cbv}, boosted dark matter in various detectors~\cite{Bringmann:2018cvk,Ema:2018bih,Lei:2020mii,An:2021qdl,PhysRevD.100.103011,Das:2021lcr,CDEX:2023wfz,Liang:2024xcx}, and crystal defects\,\cite{Budnik:2017sbu}.

In recent work~\cite{Das:2022srn}, we showed that existing low-noise quantum devices  provide strong sensitivity to dark matter with low energy deposition. Construction of such quantum devices with as low noise backgrounds as possible is currently under rapid development in fields from quantum computing to astrophysics. Quantum information is highly susceptible to decoherence, with the presence of background noise leading to faults in quantum algorithms. While there is an active field of research dedicated to solving this problem, called quantum error correction, noise is difficult to remove and currently limits the performance of quantum computers. Within astrophysics, the desire to image increasingly cold objects in the far infrared requires new cryogenic space telescopes that demand low noise equivalent power in their detectors. Broadly speaking, the most powerful new opportunities for low-threshold dark matter detection therefore are driven by advances in technology targeted at applications such as these.          

Previously, we specifically pointed out that power measurements of single quasiparticle devices provide a new detection mechanism for dark matter with low energy deposition, including both light dark matter, as well as heavier dark matter that has become bound to the Earth and is therefore not as energetic as the dark matter wind~\cite{Das:2022srn}. Such devices detect quasiparticle charges as they tunnel out of a superconducting island. Quasiparticle excitations are produced by broken Cooper pairs, which can be broken by SM sources, but also in large rates by dark matter energy deposition. As the Cooper-pair binding energy sets the energy deposition threshold, these devices lead to sensitivities to dark matter with energy depositions as low as about an meV, six orders of magnitude below the usual $\sim$keV threshold required for detection with direct detection experiments.

In this work, we consider the latest quasiparticle measurements from transmon qubits~\cite{2013NatCo...4.1913R}, and apply them to dark matter scattering predictions. A transmon qubit is a type of superconducting qubit, which is a fundamental unit of quantum information processing. It is based on a superconducting circuit consisting of a Josephson junction shunted by a capacitor. Transmon qubits have exceptionally reduced sensitivity to background noise, and therefore are ideal low-background devices for low-threshold dark matter detection.

We will show that these latest quantum measurements provide the best lab-based sensitivity to date to dark matter-nucleon scattering, for dark matter with mass lighter than about 100 MeV, and offer excellent prospects for future sensitivities. We will consider the implications for sub-component dark matter, which applies to dark matter candidates that do not comprise the entirety of the dark matter energy density observed in the Universe. Proposals to detect sub-component dark matter have become of increasing interest recently~\cite{Neufeld:2018slx,Pospelov:2020ktu, Pospelov:2019vuf, Rajendran:2020tmw,Xu:2021lmg,Budker:2021quh, McKeen:2022poo,Billard:2022cqd,Li:2022idr,Pospelov:2023mlz,McKeen:2023ztq,Ema:2024oce,Moore:2024mot}. We will calculate constraints on such a candidate stronger than any of the previous proposals, and in some cases clearly surpass spin-independent direct detection constraints on sub-component dark matter.

Our paper is organized as follows. In Sec.~\ref{sec:dminput}, we first detail the dark matter inputs relevant for our calculations. In Sec.~\ref{sec:transmon} we discuss our transmon qubit device and dark matter induced quasiparticle production and tunneling rates. In Sec.~\ref{sec:results} we discuss our new constraints on dark matter-nucleon scattering, on both the usual dark matter which explains the full energy content of the Universe, as well as sub-component dark matter candidates. We also calculate projections for future sensitivities with transmon qubits. We conclude with a summary of our findings in Sec.~\ref{sec:conclusion}.

\section{Dark Matter Density and Velocity Inputs} 
\label{sec:dminput}

There are two potential dark matter components that can be tested by direct detection experiments or our transmon qubits. As these two components have different dark matter density and velocity values, we detail them separately below.

\subsection{Dark Matter Wind from the Galactic Halo}

Dark matter which scatters with the Earth as it enters from the Galactic halo is the standard dark matter distribution tested in direct detection experiments. As such, our calculations using this dark matter distribution will be directly comparable to existing direct detection experiments, with no additional assumptions required. We will call this component ``halo dark matter".

For halo dark matter, we use a velocity distribution given by a truncated Maxwell-Boltzmann distribution,
\begin{align}
    f_\chi^{\rm halo}(\bv) = \frac{1}{N_0} e^{-(( \bv+\bv_\oplus)/\bv_{\rm halo})^2}
    \Theta(v_\mathrm{esc}^{\rm halo}-(\bv+\bv_\oplus))\,,
\label{eq:velocityh}
\end{align}
where $N_0$ normalizes the distribution,
and $\bv_{\rm halo}$ is the average dark matter velocity in the halo $\bv_{\rm halo}=230$~km/s. We incorporate the relative velocity between the Earth and the dark matter, where $|\bv_\oplus|=240\,\mathrm{km/s}$ is the Earth's velocity in the Galactic rest frame. Above $v_{\rm esc}^{\rm halo}\approx500$~km/s is the Galactic escape velocity at the Sun's position.

The local dark matter density is assumed to be the standard value of $\rho_\chi=0.4\,\mathrm{GeV/cm^{3}}$~\cite{Pato_2015}, unless otherwise stated in the case of studying sub-components (as discussed at the end of this section). 

\subsection{Thermalized Dark Matter Population Bound to the Earth}

Another dark matter component other than the usually tested halo dark matter component, is the potential thermalized dark matter population bound to the Earth. As the halo dark matter component enters the Earth and scatters, it may continue to scatter and become trapped within the Earth if it loses sufficient kinetic energy to be below the escape velocity of the Earth.  For sufficiently large dark matter-SM scattering cross sections, the dark matter can be trapped very quickly after rapidly thermalizing and entering local thermal equilibrium with the surrounding Earth matter. Under these conditions, the radial distribution of dark matter within the Earth, $n_\chi$, is described by the first-order differential equation~\cite{Leane:2022hkk}
\begin{align}
    \label{eq:radialDM}
    \frac{\nabla n_\chi}{n_\chi} + \left(\kappa+1\right) \frac{\nabla T}{T} +  \frac{m_\chi g}{T}=\frac{\Phi}{n_\chi D_{\chi N}}\frac{R_\oplus^2}{r^2}\, ,
\end{align}
where $T$ is the Earth's radial temperature profile at position $r$, $R_\oplus$ is Earth's radius, $g$ is gravitational acceleration, $D_{\chi N}\sim \lambda \,v_{\rm th}$ and $\kappa\sim-1/[2(1+m_{\chi}/m_{\rm SM})^{3/2}]$ are diffusion coefficients~\cite{Leane:2022hkk}, with $\lambda$ the dark matter mean free path, $v_{\rm th}$ the dark matter thermal velocity, and $m_\chi$ and $m_{\rm SM}$ the dark matter mass and SM target mass respectively. $\Phi$ is the incoming captured flux of dark matter particles from the Galactic halo, and is given by~\cite{Leane:2022hkk}
\begin{equation}
    \Phi = v_\chi \sqrt{\frac{8}{3 \pi}}\left[1 + \frac{3}{2}\left(\frac{v_{\rm esc}}{v_\chi}\right)^2 \right] \frac{\rho_\chi f_{\rm cap}}{m_\chi},
    \label{eq:fluxdens}
\end{equation}
where $f_{\rm cap}$ is the fraction of dark matter particles captured that pass through the Earth.

The dark matter density profile is normalized by enforcing that its volume integral equals the total number of particles expected within the Earth, i.e.~\cite{Leane:2022hkk}
\begin{align}
\label{eq:normalizationall}
    4 \pi \int_0^R r^2\, n_{\chi} \, dr = N_\chi \,,
\end{align}
where the total number of dark matter particles is given by the product of the dark matter capture rate $C$ and the lifetime of the Earth $\tau_{\rm Earth}\sim4.5$~Gyr,
\begin{equation}
    N_\chi = C\,\tau_{\rm Earth}\,.
    \label{eq:number}
\end{equation}
To obtain the capture rate $C$ for a range of dark matter masses and interaction cross sections, we use the public package \texttt{Asteria}~\cite{Leane:2023woh}. \texttt{Asteria} takes into account multiple kinematic and interaction regimes of interest, including reflection of strongly interacting light dark matter, and the relation between the fraction of dark matter particles that pass through the Earth that are captured, $f_{\rm cap}$, and the dark matter-nucleon scattering cross section, $\sigma_{\chi N}$.

As was shown in Ref.~\cite{Leane:2022hkk}, solving Eq.~(\ref{eq:radialDM}) for $n_\chi(r)$ reveals that the thermalized dark matter population has a large enhancement at the surface of the Earth, compared to the halo density value which locally for a $e.g.$ 1 GeV dark matter particle is $\sim0.4$~cm$^{-3}$ (see also the earlier Refs.~\cite{Neufeld:2018slx,Pospelov:2020ktu, Pospelov:2019vuf, Rajendran:2020tmw,Budker:2021quh, McKeen:2022poo,Billard:2022cqd}). In contrast, depending on the scattering cross section, the local dark matter density for a 1 GeV dark matter particle can be about $\sim10^{14}~$cm$^{-3}$ for the population bound to the Earth. This large enhancement in density is however somewhat offset by the fact that as this dark matter population is thermalized within the Earth, its velocity is much lower than the halo velocity. We model the thermalized dark matter velocity distribution as a truncated Maxwell-Boltzmann distribution,
\begin{align}
    f_\chi^{\rm bound}(\bv) = \frac{1}{N_0} e^{-(\bv/\bv_{\rm th})^2}
    \Theta(v_\mathrm{esc}^{\rm Earth}-v)\,,
\label{eq:velocity}
\end{align}
where $N_0$ normalizes the distribution, $v_{\rm th}^2=8T_\chi/\pi\mchi$ with $T_\chi \simeq 300\K$, and Earth's escape velocity is $v_\mathrm{esc}^{\rm Earth}\approx 11$~km/s. Dark matter being at room temperature of about $T_\chi \simeq 300\K$ is expected as dark matter is not expected to thermalize with the cold device itself, as its mean free path is much larger than the size of our device of interest. The velocity in Eq.~(\ref{eq:velocity}) requires energy detection thresholds of less than about 0.05~eV, which is not achievable with standard direct detection. As we pointed out in Ref.~\cite{Das:2022srn}, this population is already highly detectable with low-threshold quantum devices. Our transmon qubit device will provide the best existing sensitivity even for sub-component dark matter. 

An important limitation for detection of the Earth-bound thermalized dark matter distribution is a process called evaporation. In particular, when dark matter is captured inside the Earth, it receives thermal kicks from the Earth’s internal temperature. If these kicks cause dark matter to gain so much energy that it can overcome the escape velocity of the Earth, it leaves the Earth, or “evaporates”, such that there is no population remaining inside the Earth to be detected. We take the evaporation process into account by demanding that the dark matter number density is not depleted, such that total number of dark matter particles is~\cite{1987ApJ...321..560G}
\begin{align}
\label{eq:evap}
    N_\chi^{\rm tot} = \frac{C}{E}\,\left(1-e^{-E\,\tau_{\rm Earth}}\right),
\end{align}
where $C$ is the dark matter capture rate determined in \texttt{Asteria}~\cite{Leane:2023woh} as discussed above, and $E$ is the evaporation rate which is calculated following Ref.~\cite{1987ApJ...321..560G}. Here we will focus on contact interactions only, and on the case where there is no significant dark matter annihilation.

\subsection{Testing Dark Matter Sub-Components}

As well as studying dark matter which makes up all of the relic abundance of dark matter, we will also consider dark matter sub-components. Dark matter sub-components can be defined as some dark matter candidates which do not make up full the relic abundance of dark matter observed in the Universe today. That is, the candidate's contribution to the dark matter density can be expressed as some fraction $f_\chi$ of the usual $\rho_\chi=0.4\,\mathrm{GeV\, cm^{-3}}$ observed at the local position, such that the local mass density of the candidate will instead be the lower value of 
\begin{equation}
    \rho_\chi^{\rm sub}=f_\chi\times 0.4\,\mathrm{GeV\, cm^{-3}}.
    \label{eq:sub}
\end{equation}
Sub-component dark matter can be motivated from the perspective that the dark sector may be vast, with more than one particle candidate -- after all, the visible sector contains a large number of particles within the SM. There are a range of particle dark matter models considered in the literature that naturally can provide dark matter candidates in our parameter space of interest, including dark matter candidates with quantum chromodynamics (QCD) charge such as the sexaquark~\cite{PhysRevLett.38.195, Farrar:2003gh, Farrar:2005zd,Farrar:2017eqq,Moore:2024mot}. Another example explored in $e.g.$ Ref.~\cite{McKeen:2022poo} is the scenario where asymmetric dark matter is coupled to the dark photon, with predictions in previously untested parameter space to which our setup has sensitivity. In general, sub-fractions of the full dark matter density may be expected when the interaction rate is fairly large, leading to large annihilation rates in the early Universe and therefore a depletion in the relic dark matter abundance.

Detecting any potential sub-components of dark matter can be difficult. This is because testing only a very small fraction of the dark matter density requires higher-precision experiments. Proposals to detect sub-component dark matter have become of increasing interest recently~\cite{Neufeld:2018slx,Pospelov:2020ktu, Pospelov:2019vuf, Rajendran:2020tmw,Xu:2021lmg,Budker:2021quh, McKeen:2022poo,Billard:2022cqd,Li:2022idr,Pospelov:2023mlz,McKeen:2023ztq,Ema:2024oce,Moore:2024mot}. To probe sub-component dark matter, we will consider values in Eq.~(\ref{eq:sub}) of $f_\chi=1,10^{-3},10^{-6},10^{-9},10^{-12}$, to span an extreme range of sub-densities, and also to facilitate direct comparison with previous sub-component dark matter studies, which often have used these benchmarks in their projections.

Velocities of dark matter sub-components are the same for the halo or thermalized dark matter cases as relevant as discussed above in the previous subsections; only the density value is affected by being a sub-component.

\section{Dark Matter Detection with Transmon Qubits} 
\label{sec:transmon}

\subsection{Transmon Qubit Overview}

A transmon qubit is a type of superconducting qubit used in quantum computing. It is a modified version of the Cooper pair box qubit, which consists of a superconducting island connected to a superconducting reservoir by a Josephson junction. The transmon qubit is designed to have reduced sensitivity to charge noise compared to the original Cooper pair box qubit. The key feature of a transmon qubit is that it has a large shunting capacitor connected in parallel with the Josephson junction. This capacitor effectively lowers the charging energy of the qubit, reducing its sensitivity to charge noise. As a result, transmon qubits can have longer coherence times, which is essential for performing quantum computations. Transmon qubits are widely used in various quantum computing architectures, including those based on circuit quantum electrodynamics and superconducting quantum processors. They have been instrumental in many experimental demonstrations of quantum algorithms and quantum error correction codes. Their low-sensitivity to charge noise, and therefore low backgrounds for the purposes of dark matter detection, makes them ideal low-threshold dark matter detectors.

In order to investigate how quasiparticle tunneling impacts the decoherence of a transmon qubit, a study by Ref.~\cite{2013NatCo...4.1913R} devised a single-junction superconducting qubit using aluminium and examined its decoherence by tracking the rate of single-charge tunneling. By observing the relaxation rate of the qubit, they determined a quasiparticle density of $0.04\pm0.01\mum^{-3}$, characterized by a thermalized distribution~\cite{2013NatCo...4.1913R}. We previously used this device to probe dark matter scattering in Ref.~\cite{Das:2022srn}. Here we extend our investigations by testing the sensitivity of this device to subcomponents of dark matter which may interact with the device, as well as mapping out future sensitivities. We now discuss dark matter scattering rates within these devices, and how dark matter can induce quasiparticle production and tunneling that is limited by this quasiparticle density measurement.

\subsection{Calculation of Scattering Rates and Energy Deposition} 

We are interested in very low energy depositions as appropriate for light dark matter, or dark matter with low velocities. Therefore, for our regime of interest interatomic forces will be important, as the momentum transferred by dark matter $\bq$ will be comparable to the inverse size of the nuclear wavefunction in the detector crystal. In this regime, lattice vibrations or phonon excitations will provide the dark matter scattering rate, rather than the higher energy nuclear recoil regime. For our device the total scattering rate per unit target mass is\,\cite{Kahn:2021ttr,Trickle:2019nya}
\begin{align}\label{eq:rate}
    \Gamma = \frac{\pi\sigman\nchi}{\rho_T\mu^2} \int d^3v f_\chi(\bv) \int \frac{d^3q}{(2\pi)^3} \fmed^2(q) S(\bq,\oq).
\end{align}
Here, $f_\chi(\bv)$ is dark matter velocity distribution of interest as detailed in the previous section, $\rho_T$ is the target density, $\sigman$ is the dark matter-nucleon scattering cross section, $\mu$ is the reduced mass of the dark matter-nucleon system, $\fmed(q)$ is a form-factor that depends on the mediator (we assume $\fmed(q)=1$). The differential rate as a function of deposited energy $\omega$ can be written by inserting a delta function,
\begin{align}\label{eq:rate_spectrum}
    \frac{d\Gamma}{d\omega} = \frac{\pi\sigman\nchi}{\rho_T\mu^2} &\int d^3v\, f_\chi(\bv) \\ &\times\int \frac{d^3q}{(2\pi)^3}\, \fmed^2(q)\, S(\bq,\omega)\, \delta(\omega-\oq)\,,\nonumber
\end{align}
where the dark matter energy deposition is given by 
\begin{equation}
    \oq = \bq\cdot\bv - \frac{q^2}{2\mchi}.
\end{equation}
In Eq.~(\ref{eq:rate}), $S(\bq,\oq)$ is the dynamic structure factor containing the detector response to dark matter scattering and depends on the crystal structure of the target material, which is given by
\begin{align}
\label{eq:structure_factor}
    S(\bq,\omega_{\bq}) &\approx \dfrac{2\pi}{V_c} \sum_df_d^2\, e^{-2W_d(\bq)} \sum_n \left(\dfrac{q^2}{2m_d}\right)^n 
     \dfrac{1}{n!}\\
     &\times\left(\prod_{i=1}^n \int d\omega_i \dfrac{D_d(\omega_i)}{\omega_i}\right) \delta\left(\omega-\sum_i\omega_i\right)\,,\nonumber
\end{align}
\begin{figure}[t]
    \centering
\includegraphics[width=0.485\textwidth]{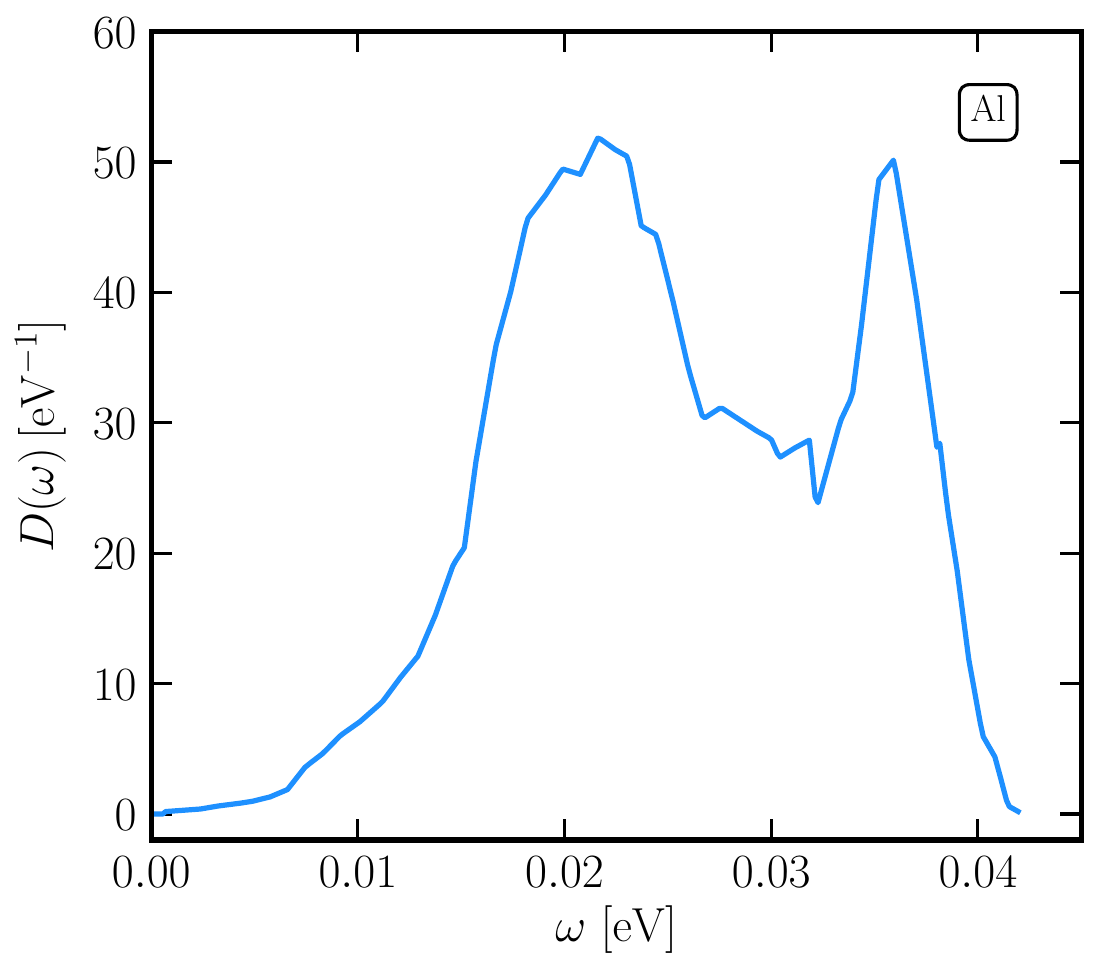}
    \caption{Phonon density of state $D(\omega)$ for our target aluminium as a function of energy $\omega$.}
    \label{fig:dos}
\end{figure}
in the large momentum transfer regime\,\cite{Campbell-Deem:2019hdx}. Here, $V_c$ is the primitive cell volume, $n$ is the number of phonons excited, $m_d$ is the mass of the target atoms. The average phonon energy $\bar{\omega}_d$, determined by the density of state of the material, determines the typical number of phonons excited, 
\begin{equation}
    n\approx \frac{q^2}{2m_d\bar{\omega}_d}.
\end{equation} 
Here, as we have an aluminium target, $\bar{\omega}_d = 25.5$~meV and $m_d = 27\GeV$. As an example benchmark, for thermalized dark matter with 1 GeV mass and a maximum momentum transfer of $q = 20$ keV, the typical phonon number is therefore $n = 0.29$.

Figure~\ref{fig:dos} shows $D(\omega)$, the  phonon density of state of the single atom in the primitive cell, for our target aluminium. The phonon density of state is important for understanding at what energy resonant energy transfer occurs, and therefore at what energies are there especially large scattering rates. The peaks shown in the aluminium phonon density of state are from the transverse and longitudinal acoustic phonon branches. Aluminium has a face-centered-cubic crystal structure with only one atom in its primitive cell, which implies that it only has three acoustic phonon branches.

In Eq.~(\ref{eq:structure_factor}), we take $f_d=A_d$ as the coupling for spin-independent interactions, where the scattering benefits from nuclear coherence. Above the Debye-Waller function is given by
\begin{align}\label{eq:debye-waller}
    W_d(\bq) = \dfrac{q^2}{4m_d} \int d\omega\, \dfrac{D_d(\omega)}{\omega}\,.
\end{align}
To implement scattering rates we use the publicly available code \texttt{DarkELF}~\cite{Knapen:2021bwg,Campbell-Deem:2022fqm}, with some additional modifications. To allow for tests of both halo dark matter and Earth-bound dark matter, we update the local dark matter density and velocity inputs as detailed in the previous section. We also modify the code to apply to aluminium, which is our target material. As discussed above, aluminium has only one atom in the primitive cell. Therefore, we modify \texttt{DarkELF} to consider only one atom in the primitive cell, rather than for two atoms per primitive cell as per the default \texttt{DarkELF} setup.

\subsection{Dark Matter Induced Quasiparticle Density}

In superconducting metals, the low temperature allows electrons to form Cooper pairs. These Cooper pairs are bound via long-range interactions with phonons. When low-energy dark matter scatters with the superconducting aluminium film as described in the previous subsection, its kinetic energy can be deposited in the form of phonons. These phonons can break the Cooper pairs if they have energy exceeding the Cooper pair binding energy of $\sim$meV. This leads to an excess of quasiparticles in the metal. These excess quasiparticles may tunnel out of the superconducting island, and the total number of quasiparticles present at a given time is sensitive to the rate in which they are produced, are trapped, or recombine. This can be determined using the mean field results~\cite{bespalov}
\begin{align}
    \dfrac{d\nqp}{dt} &= -\Gamma_R - \Gamma_T + \Gamma_G \nonumber\\
    &\approx -\Bar{\Gamma}\nqp^2 - \Bar{\Gamma}_T\nqp + \Gamma_G\,,
    \label{eq:meanf}
\end{align}
where $\nqp$ is the steady-state quasiparticle density, and $\Gamma_R, \Gamma_T, \Gamma_G$ are the recombination, trapping, and generation rates, respectively. The quasiparticle generation rate is related to the steady state power density injected by dark matter $P_{\rm DM}$ as
\begin{align}
    \Gamma_G &= \frac{\epsilon_{\rm qp}}{2\Delta}\int d\omega~ \omega\,\frac{d\Gamma}{d\omega}\\
    &=\frac{P_{\rm DM}}{2\Delta},
\end{align}
where $\Delta$ is the superconducting energy gap, $\epsilon_{\rm qp}$ is the quasiparticle generation efficiency. We assume a quasiparticle generation efficiency of 60\% ($\epsilon_{\rm qp}=0.6$)~\cite{Kaplan,Hochberg:2021ymx}, and an energy gap $\Delta \simeq 340\mueV$ for aluminium. 

In equilibrium, $i.e.$ when the quasiparticle density is constant in time, Eq.~(\ref{eq:meanf}) then simplifies to
\begin{equation}
\frac{P_{\rm DM}}{2\Delta} = \Bar{\Gamma}\nqp^2 + \Bar{\Gamma}_T\nqp \,.
\end{equation}
Under the assumption of no quasiparticle trapping, the expression is further simplified, and the quasiparticle density is then related to the dark matter power injection as
\begin{equation}
    \nqp=\sqrt\frac{P_{\rm DM}}{2\Delta\, \Bar{\Gamma}},
\end{equation}
where $\bar{\Gamma}=40\,\mathrm{s^{-1}\mu m^3}$ for aluminium\,\cite{bespalov}. The quasiparticle generation rate  $\bar{\Gamma}$ has some associated uncertainty due to the limitations of the quasiparticle diffusion model, and so when making plots we will take $\overline{\Gamma} = 4$ or $400\,\mathrm{s^{-1}}\mum^3$ for conservative and optimistic values respectively.
\begin{figure}[t]
    \centering
\includegraphics[width=\columnwidth]{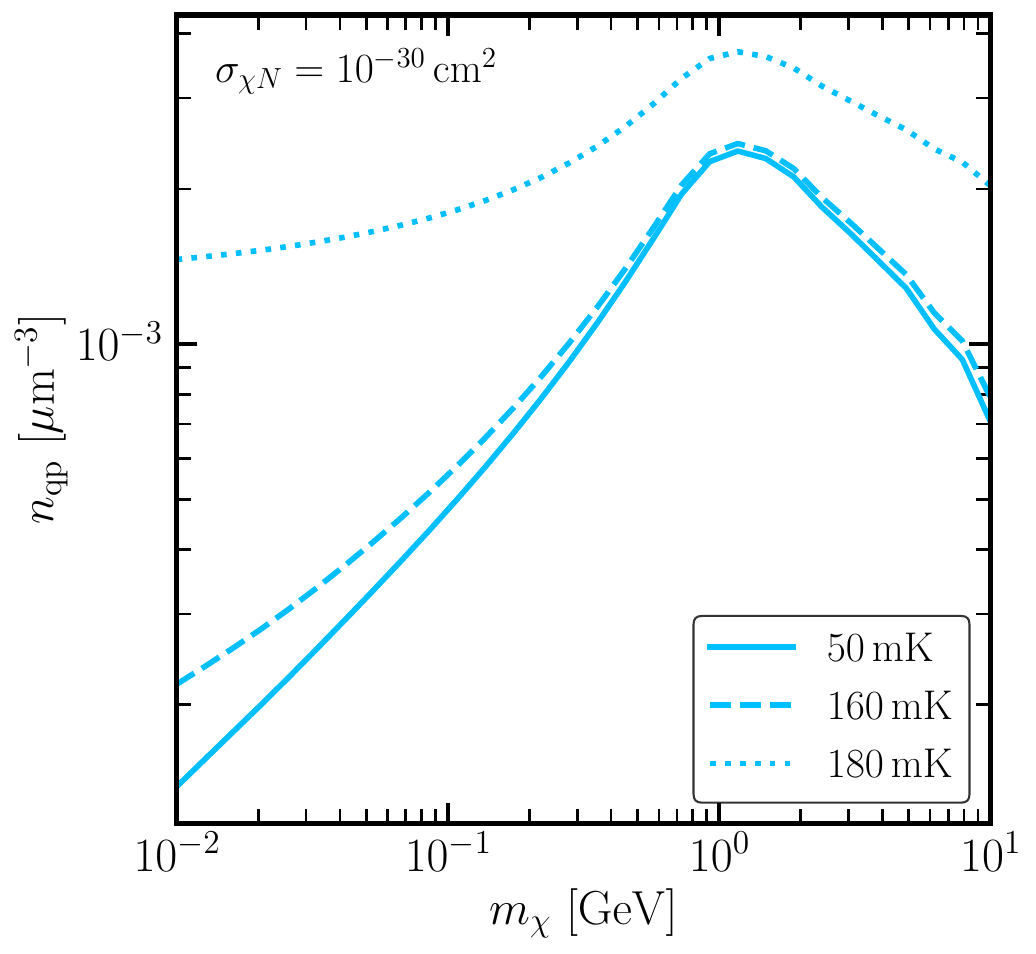}
    \caption{Dark matter induced quasiparticle density $n_{\rm qp}$ in aluminium for three benchmark temperatures $T=50, 160, 180\,$mK and a benchmark dark matter-nucleon scattering cross section of $\sigman=10^{-30}\cmsq$. Plot assumes halo dark matter.}
    \label{fig:qp_density}
\end{figure}

The steady-state density induced by dark matter is therefore
\begin{equation}\label{eq:qp_density}
    \nqp \approx \left(\frac{P_{\rm DM}}{3.6\times 10^{-21}\mathrm{W}}\right)^{1/2}\mum^{-3}.
\end{equation}
For our transmon qubit, this implies that the upper limit of any residual power injection is $3.92\times 10^{-24}\Wmum$, for dark matter or any other additional source. Given the scattering formalism discussed in the previous subsection, this then translates directly into a constraint on the dark matter scattering rate.
Note that by considering this as an upper limit, we are being conservative. This is because the source of the quasiparticle density measured is not known, and as we pointed out in Ref.~\cite{Das:2022srn}, it may already include a possible positive dark matter signal.

The total quasiparticle density at a temperature $T$ is given by the sum of the dark matter-induced component and the thermal population,
\begin{align}
    \nqp^\mathrm{tot} = \nqp + 2\nu_0\Delta \sqrt{\dfrac{2\pi k_BT}{\Delta}}\exp\left(-\dfrac{\Delta}{k_BT}\right)\,.
\end{align}
Here, $\nu_0=1.2\times10^4\mum^{-3}\,\mathrm{\mu eV}^{-1}$ is the Cooper pair density of state at the Fermi level in aluminium.

Figure~\ref{fig:qp_density} shows this quasiparticle density in aluminium for three benchmark temperatures $T=50, 160, 180\,$mK and a benchmark dark matter-nucleon scattering cross section of $\sigman=10^{-30}\cmsq$. These choices are motivated as the 180 mK line is effectively the setup of Ref.~\cite{Connolly:2023gww}, except recombination limited, and 50 mK is chosen as another benchmark as it corresponds to approaching the limit of what can be realistically achieved. 160 mK is an arbitrary intermediate value. This plot assumes the dark matter inputs from the incoming Galactic halo dark matter; quasiparticle densities can be higher for the Earth-bound thermalized dark matter.

\subsection{Other Single Quasiparticle Devices}

In our previous work~\cite{Das:2022srn}, we considered two other low quasiparticle density devices other than transmon qubits; a low-noise bolometer~\cite{QCD}, and SuperCDMS-CPD~\cite{Fink_2021}, whose power measurements had not previously been considered for a dark matter search. We now briefly comment on these devices, which can produce weaker but supporting bounds in addition to our transmon qubit.

Ref.~\cite{QCD} developed a quantum capacitance detector where photon-produced free electrons in a superconductor tunnel into a small capacitive island. This setup is embedded in a resonant circuit, and therefore can be referred to as a ``quantum resonator". This quantum resonator measured excess power of $4\times 10^{-20}\W$~\cite{QCD}, making it the most sensitive existing far-infrared detector. Given the power measurement of this device, it is a few orders of magnitude less sensitive than our transmon qubit. 

SuperCDMS detectors are transition-edge sensors (TES). The lowest bias power measurement is from SuperCDMS-CPD~\cite{Fink_2021}, which is a TES is coupled to a large silicon absorber \cite{Ren_2021}. SuperCMDS-CPD measured an excess power of 6\,pW was measured in the phonon sensor arrays, which is an excess substrate power of $10^{-24}\Wmum$. This therefore produces weaker or comparable bounds to our transmon qubit, depending on the accuracy of the quasiparticle generation rate.

Both of these measurements support our constraints, especially SuperCDMS-CPD's bias power measurements. Note that for both of these devices, we conservatively only considered dark matter detection from the superconducting films, rather than the substrate, see Ref.~\cite{Das:2022srn} for extended discussion on this point.

\section{New Dark Matter Parameter Space Sensitivities} 
\label{sec:results}

\begin{figure}[b]
    \centering
    \includegraphics[width=\columnwidth]{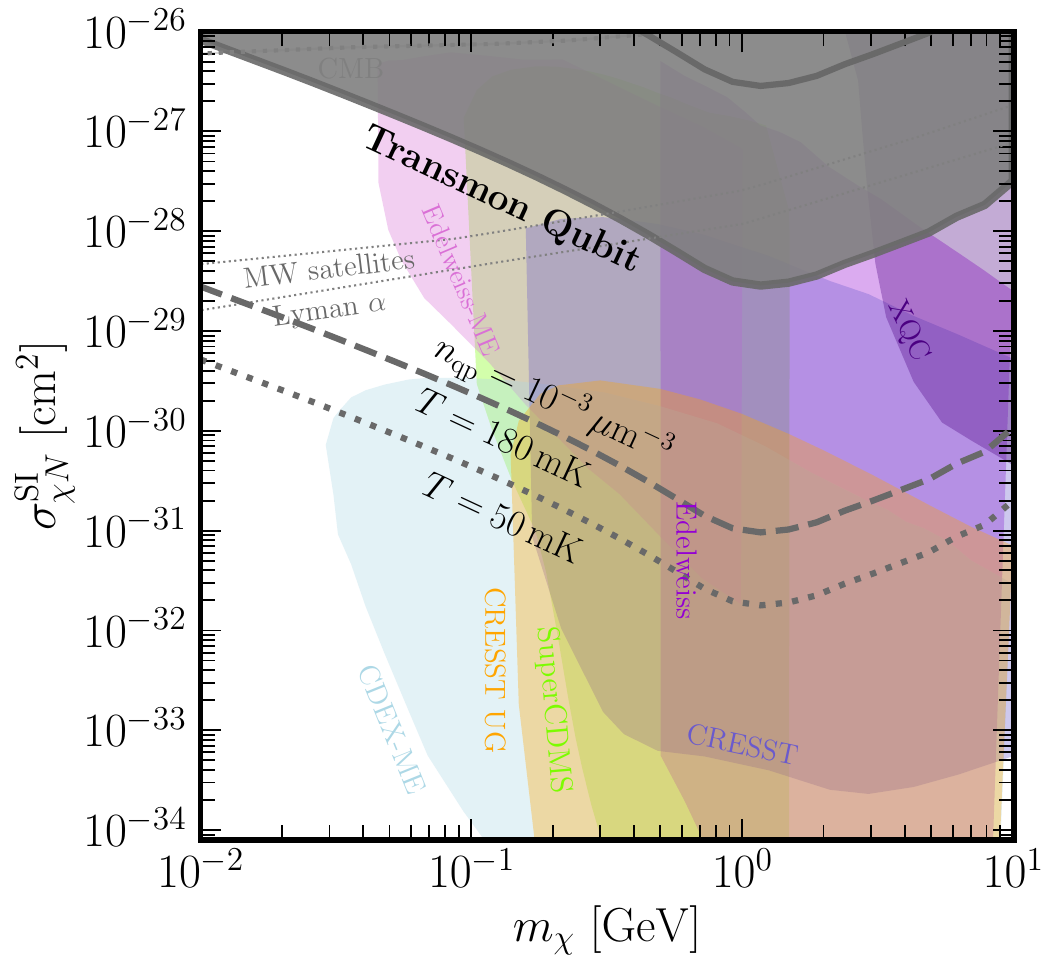}
    \caption{Limit on the dark matter-nucleon scattering cross section (gray-shaded) for incoming Galactic dark matter, with the two lines being conservative and optimistic values of the quasiparticle generation rate. We show two projections (black dashed and dotted) with quasiparticle density $n_\mathrm{qp}=10^{-3}\mum^{-3}$ for two temperatures as labeled, see text for motivation.}
    \label{fig:limits_halo}
\end{figure}

Figure~\ref{fig:limits_halo} shows the bounds and sensitivities we derive for spin-independent scattering using a transmon qubit, using the incoming Galactic dark matter, under the assumption the dark matter makes up the full relic abundance ($i.e.$ $f_\chi=1$). This is in direct one to one correspondence with the usual direct detection bounds, that is, all lines here have about the same assumptions of the local dark matter density and velocity.  The limits in Fig.~\ref{fig:limits_halo} can be simply rescaled linearly in $f_\chi$ for subcomponent dark matter. We see that our bounds on halo dark matter are stronger than lab-based experiments below about 100 MeV, as we already had shown in Ref.~\cite{Das:2022srn}. We also show new projected sensitivities for a few benchmark scenarios, assuming improved future single quasiparticle devices. We have taken a quasiparticle density about an order of magnitude smaller than our existing transmon qubit, as labeled as the dashed line with $n_{\rm qp}=10^{-3}\mum^{-3}$, at two quasiparticle temperatures of 180 mK and 50 mK as labeled. As noted in the previous section, these choices are motivated as the 180 mK line is effectively the setup of Ref.~\cite{Connolly:2023gww}, except recombination limited, and 50 mK is chosen as another benchmark as it corresponds to approaching the limit of what can be realistically achieved in the near future. Note that all estimates assume recombination-limited quasiparticle density modeling. Active work to understand sources of quasiparticle poisoning for qubit error mitigation, both from qubit drive~\cite{Amin:2024pag} and radiation~\cite{Fink:2023tvb,Yelton:2024tqo}, will contribute to even more stringent sensitivities than the projections shown, as quasiparticle backgrounds are reduced further.

In Fig.~\ref{fig:limits_halo} we also compare with existing limits, including those from astrophysical systems such as Milky Way satellites~\cite{DES:2020fxi}, Lyman-alpha~\cite{Rogers:2021byl}, and the Cosmic Microwave Background (CMB)~\cite{Xu:2018efh}. There are also lab experiments overlapping with part of our parameter space, namely CRESST~\cite{CRESST:2017ues}, SuperCDMS~\cite{SuperCDMS:2020aus}, Edelweiss~\cite{EDELWEISS:2019vjv}, CDEX\,\cite{CDEX:2021cll}, XQC~\cite{Mahdawi:2018euy}, and CRESST UG\,\cite{CRESST:2017cdd}. 
Note that for cross sections above about $10^{-30}$~cm$^2$, the Born approximation breaks down, and the nuclear coherence across different detector materials is not well defined without using a particle dark matter model, see $e.g.$ Refs.~\cite{Digman:2019wdm,Xu:2020qjk}. 

 \begin{figure*}[]
    \centering
    \includegraphics[width=0.45\textwidth]{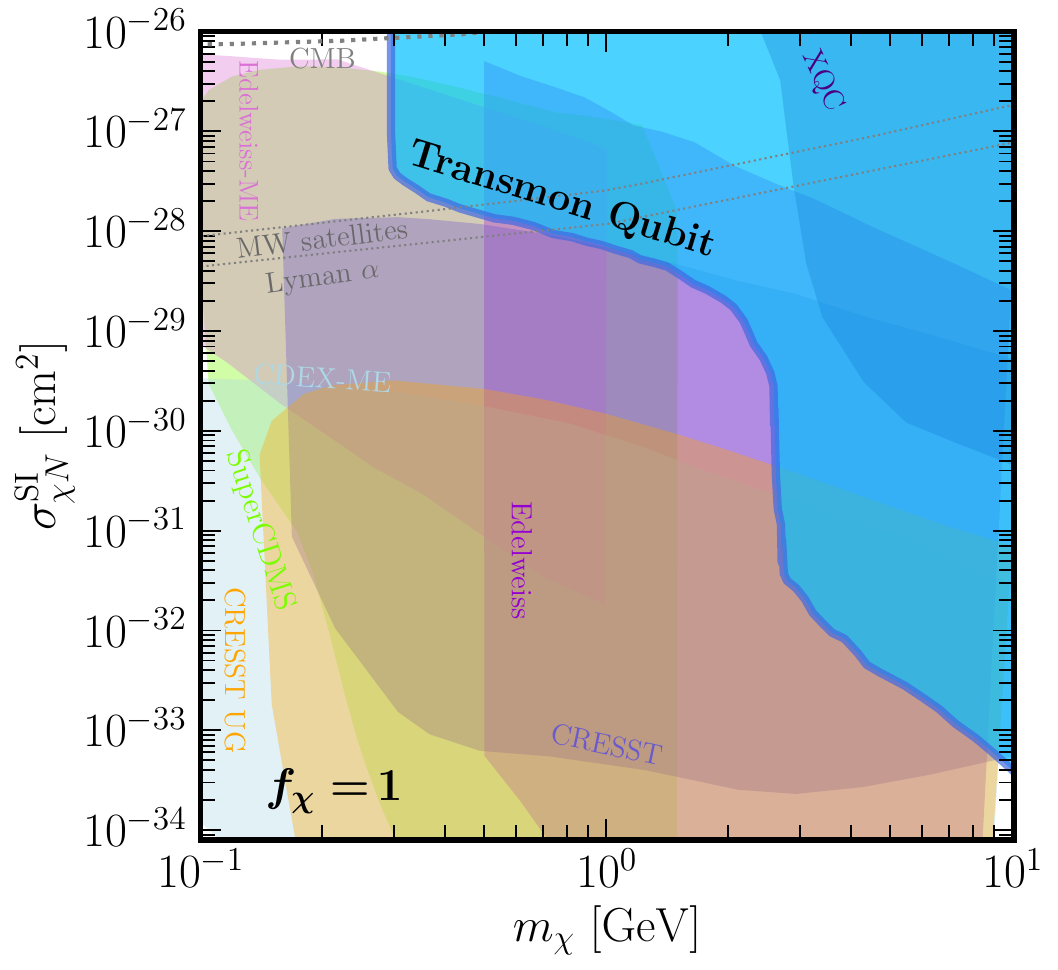}
    \includegraphics[width=0.45\textwidth]{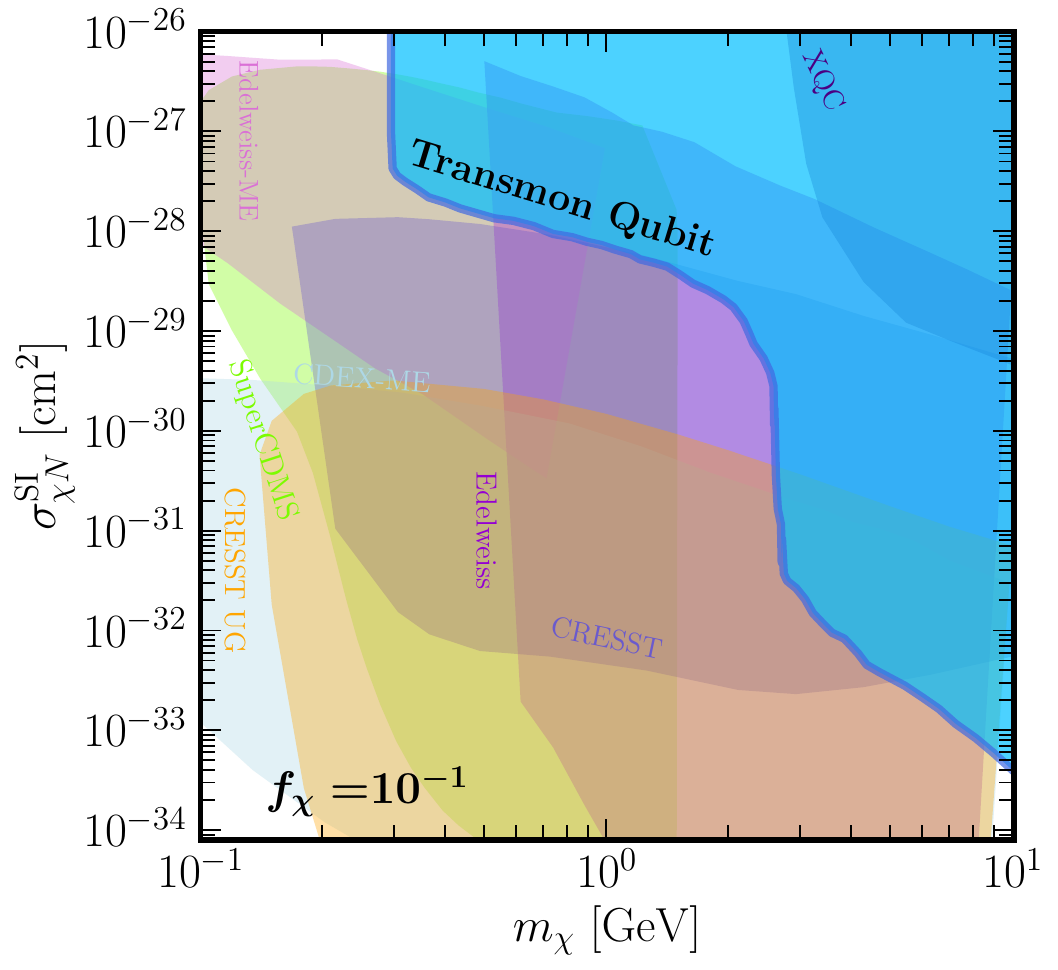}\\
    \includegraphics[width=0.45\textwidth]{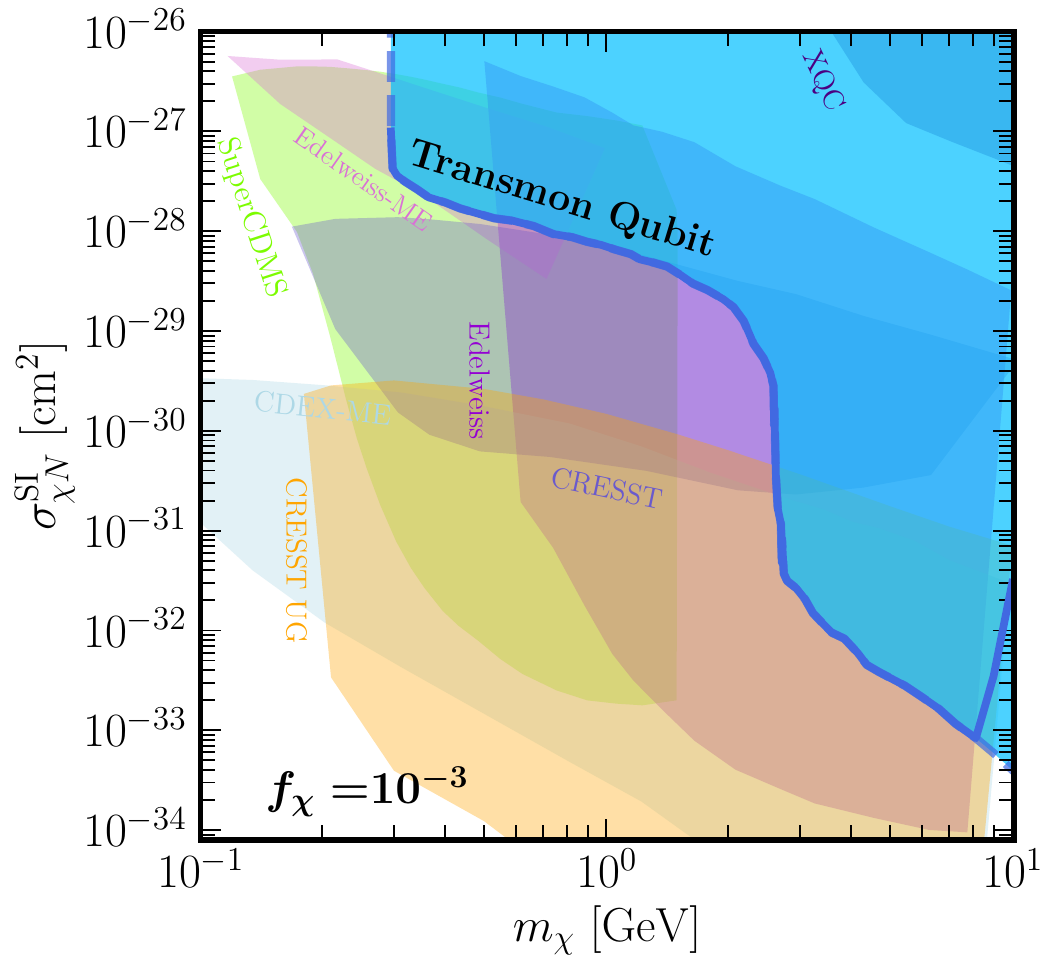}
    \includegraphics[width=0.45\textwidth]{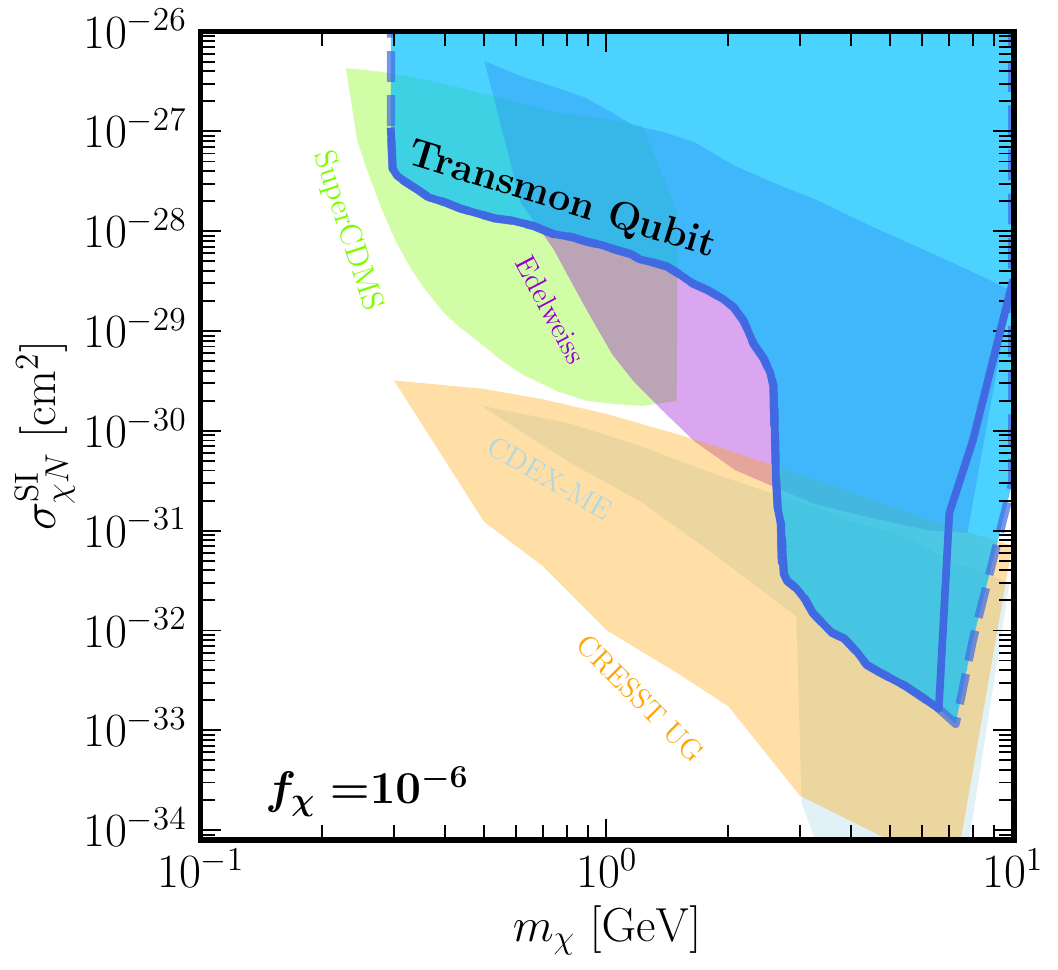}\\
    \includegraphics[width=0.45\textwidth]{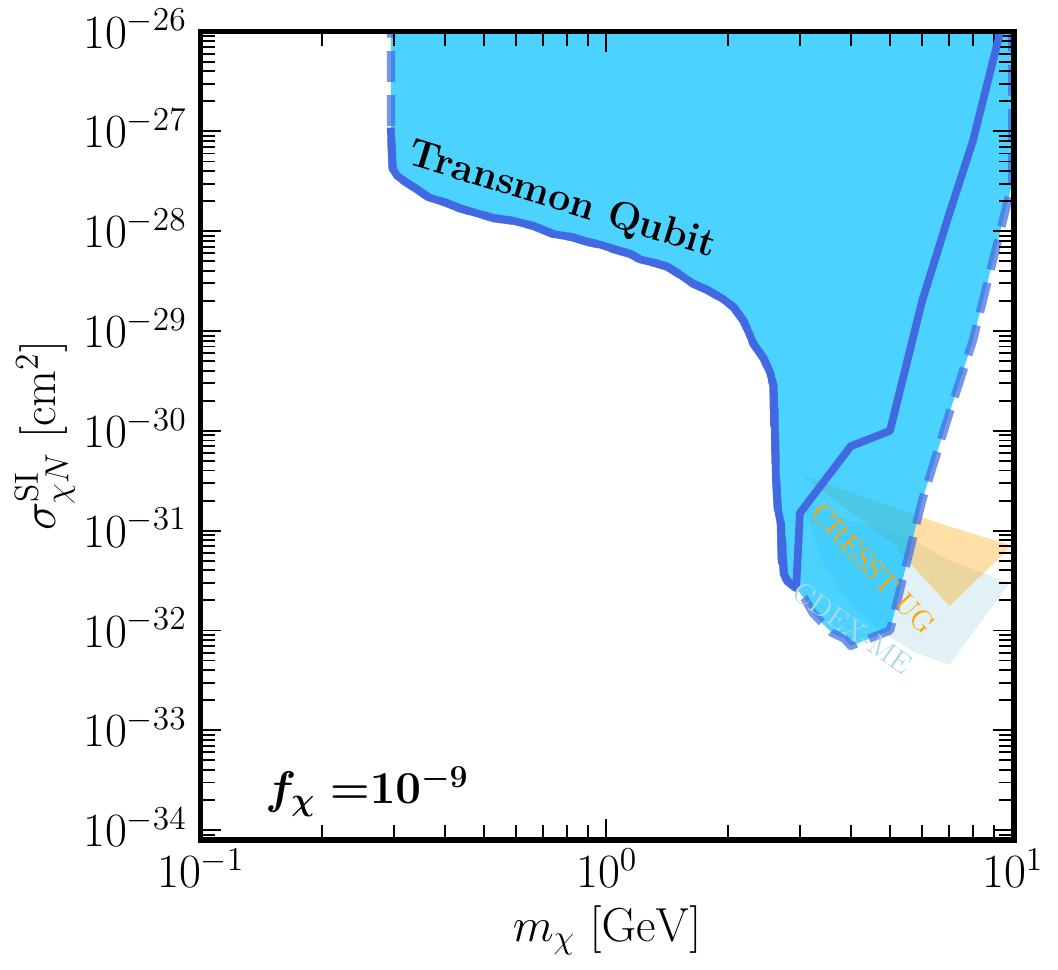}
    \includegraphics[width=0.45\textwidth]{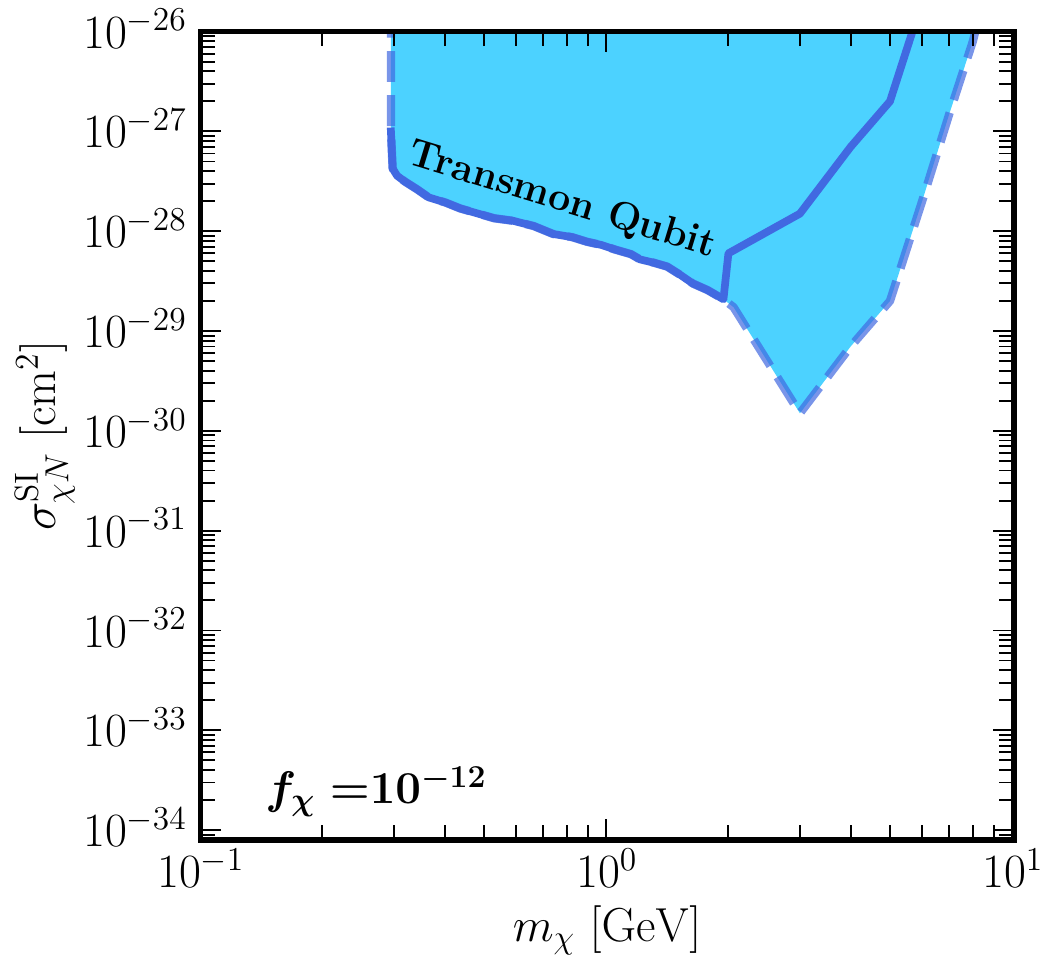}\\
    \caption{New constraints on the dark matter-nucleon spin-independent scattering cross section $\sigma_{\chi N}^{\rm SI}$ as a function of dark matter mass $m_\chi$, using transmon qubit measurements for the Earth-bound dark matter population, for varying fractions of the dark matter density $f_\chi$ as labeled. Solid line is the conservative bound, dashed is the optimistic bound (see text for details). Complementary limits are also shown as labeled.}
    \label{fig:limits_therm}
\end{figure*}

Figure~\ref{fig:limits_therm} shows our new limits on the dark matter-nucleon spin-independent scattering cross section as a function of dark matter mass, using transmon qubit measurements for the Earth-bound dark matter population, for varying fractions of the dark matter density $f_\chi$ as labeled. The dashed line corresponds to the optimistic bound based on quasiparticle generation rate uncertainties, and the solid line is conservative. We see that while the $f_\chi=1$ case (dark matter candidate corresponds to the full dark matter density) overlaps most with existing experiments, as soon as this density is decreased other experiments weaken or disappear, while our bounds are not largely changed in comparison. 

In Fig.~\ref{fig:limits_therm}, note that we linearly rescale the complementary direct detection limits only from below; the ceilings of the limits also decrease but non-linearly with decreasing dark matter fraction, and require a dedicated calculation~\cite{Cappiello:2023hza}. The reason the ceilings scale non-linearly with decreasing dark matter fraction, is that the dark matter velocity is also relevant when the scattering limit is near the detection threshold~\cite{Cappiello:2023hza}. As only CRESST has had its limit ceilings recast for subfractions~\cite{Cappiello:2023hza}, we adopt the recast CRESST bounds for both the lower and upper ends of the bounds, but only rescale the other direct detection constraints from below. From the point of view of how much new parameter space our transmon qubit probes, this is a conservative choice, although the ceiling limit change with decreasing dark matter subfractions in any case is not very large~\cite{Cappiello:2023hza}. In the case where $f_\chi=0.1$, the astrophysical bounds weaken or disappear. This is because limits relying on structure formation require the dark matter candidate to make up most of the dark matter content, and as soon as dark matter is a smaller component, it has a negligible effect on structure. 

In Fig~\ref{fig:limits_therm}, we show an extreme range of sub-fractions of dark matter, to facilitate direct comparison to the same values chosen in the literature, but also to demonstrate the strong constraining power of transmon qubits. We produce limits stronger than any other existing study on the Earth-bound thermalized population, and even produce exclusions which outperform the projections of most of the proposals to detect this population. We see that even for subfractions that are as tiny as $10^{-12}$, we still have sensitivity to dark matter parameters. On the other hand, existing direct detection experiments completely lose sensitivity for sub-components below about $10^{-9}$, as shown in Fig.~\ref{fig:limits_therm}.

Notice that all the thermalized dark matter bounds in Fig.~\ref{fig:limits_therm} are truncated by the same S-shaped curve. This line is not due to inherent sensitivity in the transmon qubit, but rather due to evaporation. As discussed earlier, when dark matter is captured inside the Earth, it receives thermal kicks from the Earth's internal temperature. If these kicks cause dark matter to gain so much energy that it can overcome the escape velocity of the Earth, it leaves the Earth, such that there is no population remaining inside the Earth to be detected. This is actually our key limiting factor across most of the parameter space for the Earth-bound dark matter. This evaporation line is obtained for non-annihilating dark matter following the original treatment of Ref.~\cite{1990ApJ...356..302G} (see Eq.~(\ref{eq:evap})); note that other works studying the Earth-bound population have often instead used the surface of last scattering, which is less accurate at lower scattering cross sections~\cite{1990ApJ...356..302G}. Any difference in the location of the evaporation-limited truncation of our bounds compared to other works studying this Earth-bound population should be applied equally; $i.e.$ our transmon qubit sensitivities themselves are not actually weaker, instead generally our treatment of evaporation is more conservative, and if we adopt the surface of last scattering treatment as per other Earth-bound works, we would obtain the same limit truncation as other works.

\begin{figure}[t]
    \centering
    \includegraphics[width=\columnwidth]{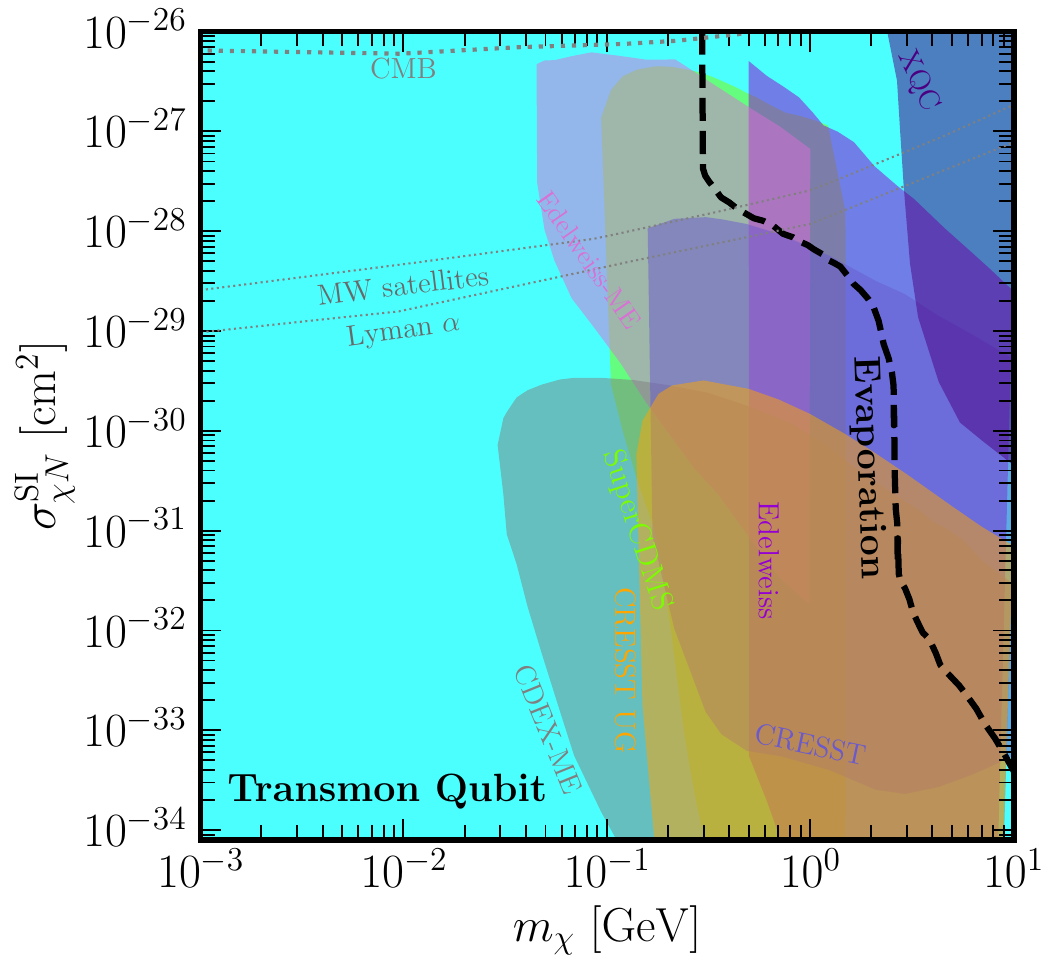}
    \caption{Constraints on the dark matter-nucleon scattering cross section as a function of dark matter mass, using transmon qubit data assuming that the thermalized dark matter does not evaporate (cyan shaded, full plot range). The evaporation line for dark matter in the Earth is shown for comparison as the dashed line, and complementary constraints are also shown.}
    \label{fig:no_evap}
\end{figure}

Figure~\ref{fig:no_evap} shows the strength of the transmon qubit constraints on the thermalized dark matter population (with $f_\chi=1$), assuming no evaporation. Here this underscores the point discussed in the previous paragraph, that our results are largely evaporation limited, not detector sensitivity limited. The dashed line for comparison shows where evaporation rapidly removes the dark matter population from the Earth (under the assumption of contact interactions); any other treatment of dark matter evaporation may simply cut our plot with where evaporation becomes relevant for the given scenario. One scenario in which this is interesting, is where evaporation is inhibited due to the presence of long-range forces~\cite{Acevedo:2023owd}. However it is important to note in that case that long-range forces which serve to avoid evaporation generally also alter the dark matter distribution inside the Earth. In some cases this can be significant, as avoiding evaporation requires significantly reinforcing gravity, and gravity is a key input for the dark matter distribution. We have not altered the distribution inside the Earth in this plot, as the precise way this happens will depend on the particle model. Instead here we aim to make the point that (i) our detector has strong sensitivity outside the evaporation range, and evaporation is largely what limits our constraints, as well as (ii) that our transmon qubit may provide strong sensitivities to particle dark matter models which do not evaporate at these masses. 

Note that in Fig.~\ref{fig:no_evap} we have restricted the cross section range to be the same as the other figures, as this is the regime where the dark matter mean free path is much shorter than the size of the Earth. The transmon qubit certainly has sensitivity at even lower cross sections, but for simplicity and the point we want to make here, we do not study the long-mean free path regime.

Fig.~\ref{fig:no_evap} also helps make sense of the shape of the halo dark matter limits in Fig.~\ref{fig:limits_halo} compared to the thermalized dark matter limits: the halo limits do not care about evaporation, because even dark matter which will later evaporate does first enter the Earth at the detector position, which is the stage in which the limit is set for regular halo dark matter. The halo limits therefore extend to lower dark matter masses than the thermalized dark matter limits, and do not feature the S-like shape at their boundary. It also explains why the Earth-bound thermalized dark matter limits do not weaken linearly with the dark matter density sub-fraction value; it is not until the sub-fraction causes the sensitivity to weaken above the evaporation threshold that the linear scaling with dark matter density is revealed, as is seen in the bottom two panels of Fig.~\ref{fig:limits_therm}.

\section{Summary and Outlook}
\label{sec:conclusion}

Low-energy dark matter interactions, such as those arising from light dark matter or dark matter thermalized and bound to the Earth, require new techniques for detection. In this work we have investigated the sensitivity of a transmon qubit with world-record low quasiparticle density to low-energy dark matter interactions. This transmon qubit is advantageous over traditional direct detection experiments, as its energy deposition detection threshold is as low as about an meV, six orders of magnitude lower than standard direct detection setups. The meV threshold is set by the Cooper-pair binding energy in the superconducting island, as when the Cooper pairs are broken by dark matter interactions, they release detectable quasiparticles. The transmon qubit can be therefore substantially more sensitive than direct detection to low-energy dark matter interactions.

We investigated current bounds and future sensitivities for two low-energy dark matter components: light dark matter incoming from the Galactic halo, as well as the dark matter distribution thermalized and bound to the Earth. For the thermalized population, there have recently been a growing number of proposals to detect it, especially if dark matter which scatters with the detector makes up some sub-component of the full dark matter relic density. We set new constraints on dark matter sub-fractions as low as $10^{-12}$, to which no existing experiment is currently sensitive. For many other subfractions, our transmon qubit provides the strongest constraints to date on thermalized dark matter. For the thermalized population, we also showed how the sensitivity depends on the evaporation threshold, with the main limitation of our sensitivity arising from evaporation rather than the transmon qubit's sensitivity itself. We showed that these devices may be fruitful moving forward for detecting dark matter particle models that do not evaporate at very light dark matter masses.

For the Galactic halo dark matter population, transmon qubits provide constraints on dark matter-nucleon scattering stronger than direct detection for dark matter masses below about 100 MeV. We showed the future promise for transmon qubits or any low-quasiparticle density device to detect light dark matter, considering both lower quasiparticle densities, as well as devices cooled to lower temperatures. We expect that low quasiparticle density devices can reach cross sections as low as about $10^{-31}$~cm$^2$ for dark matter masses below about 100 MeV for Galactic halo dark matter, and many orders of magnitude below this for thermalized dark matter. Active work to understand sources of quasiparticle poisoning for qubit error mitigation, both from qubit drive~\cite{Amin:2024pag} and radiation~\cite{Fink:2023tvb,Yelton:2024tqo} will lead to reduced quasiparticle backgrounds, and even better sensitivities in the future as quasiparticle backgrounds are reduced further. Overall, new low-quasiparticle density devices as they are actively created for applications such as quantum computing, infrared telescopes, or radiation detectors, offer substantial potential to probe dark matter with low energy deposition, and are a promising path forward to detect light dark matter.

\section*{Acknowledgments} 

We thank Chris Cappiello and Tom Connolly for helpful discussions. AD was supported by
Grant Korea NRF-2019R1C1C1010050. NK and RKL are supported by the U.S. Department of Energy under Contract DE-AC02-76SF00515. NK is also supported by the US. Department of Energy Early Career Research Program (ECRP) under FWP 100872.

\bibliography{DM_noise}
\bibliographystyle{apsrev4-2}

\end{document}